\documentclass{article}
\usepackage[utf8]{inputenc}
\usepackage[]{graphicx}
\usepackage[]{xcolor}
\usepackage{alltt}
\usepackage[left=2cm,right=2cm, top = 2.2cm, bottom = 2.2cm]{geometry}
\usepackage{amsmath}
\usepackage{amssymb}
\RequirePackage[authoryear]{natbib}
\RequirePackage[colorlinks,citecolor=blue,urlcolor=blue]{hyperref}
\RequirePackage{booktabs}
\usepackage{authblk}

\begin{document}

\title{Direction Augmentation in the Evaluation of Armed Conflict Predictions}

\author[1, 2]{Johannes Bracher} 
\author[1]{Lotta R\"uter}
\author[1]{Fabian Kr\"uger}
\author[1, 2]{Sebastian Lerch}
\author[1, 2]{Melanie Schienle}

\renewcommand\Affilfont{\fontsize{9}{10.8}\itshape}
\affil[1]{Chair of Statistical Methods and Econometrics, Karlsruhe Institute of Technology (KIT), Karlsruhe, Germany}
\affil[2]{Computational Statistics Group, Heidelberg Institute for Theoretical Studies (HITS), Heidelberg, Germany}

\maketitle

\begin{center}
Contact: \url{johannes.bracher@kit.edu}
\end{center}

\begin{abstract}
In many forecasting settings, there is a specific interest in predicting the sign of an outcome variable correctly in addition to its magnitude. For instance, when forecasting armed conflicts, positive and negative log-changes in monthly fatalities represent escalation and de-escalation, respectively, and have very different implications. In the ViEWS forecasting challenge, a prediction competition on state-based violence, a novel evaluation score called targeted absolute deviation with direction augmentation (TADDA) has therefore been suggested, which accounts for both for the sign and magnitude of log-changes. While it has a straightforward intuitive motivation, the empirical results of the challenge show that a no-change model always predicting a log-change of zero outperforms all submitted forecasting models under the TADDA score. We provide a statistical explanation for this phenomenon. Analyzing the properties of TADDA, we find that in order to achieve good scores, forecasters often have an incentive to predict no or only modest log-changes. In particular, there is often an incentive to report conservative point predictions considerably closer to zero than the forecaster's actual predictive  median or mean. In an empirical application, we demonstrate that a no-change model can be improved upon by tailoring predictions to the particularities of the TADDA score. We conclude by outlining some alternative scoring concepts.

\end{abstract}

\section{Introduction: The ViEWS competition and forecast evaluation}
\label{sec:introduction}

The ViEWS\footnote{ViEWS stands for \textit{Violence Early Warning System}, a publicly available ensemble approach for predicting monthly indicators of violence on a global scale, see \cite{Hegre2019} for more details.} prediction competition \citep{Hegre2022, Vesco2022} represents a major effort to improve forecasting capacities in the field of armed conflicts studies. It provides a valuable opportunity to compare various statistical and machine learning methods, as well as combined ensemble forecasts.

The prediction target defined for the competition was the log-change in monthly fatalities due to state-based violence in a given region. In slightly modified notation, the target for a forecast referring to month $t$ and issued at the initial time $t - s$ with a lead time of $s$ months was
$$
Y_{t, s} = \log(X_{t} + 1) - \log(X_{t - s} + 1),
$$
where $X_{t}$ is the number of fatalities in month $t$. A value of $Y_{t, s} = 0$ thus corresponds to no change between months $t - s$ and $t$, while negative values imply a decrease and positive values an increase in fatalities. Participants of the competition were asked to provide point forecasts for different initial and lead times. The geographic focus of the challenge was Africa, and forecasts could be issued at the country level and for sub-national units defined via a grid.

In \citeauthor{Vesco2022} (\citeyear{Vesco2022}; Figure 6) the collected forecasts were assessed via a total of 13 different statistical evaluation scores. These aimed to cover different desirable properties of point forecasts, including their accuracy and positive impact on ensemble forecasts. The main accuracy metrics were the following: (i) the mean squared error (MSE), which had been pre-specified in the announcement of the challenge, and (ii) the newly introduced \textit{Targeted Absolute Distance with Direction Augmentation} (TADDA). TADDA is designed as a ``metric specifically tailored to evaluate predictions of changes in fatalities, as it accounts for both the sign and the magnitude of the predictions versus the actual change'' \cite[p.~542]{Hegre2022}. To this end, TADDA combines a measure of distance between the forecast and the observation and a term penalizing forecasts with a different sign than the observed value. See the next section for the exact definition.

The independent scoring committee (SC) convened to evaluate forecasts ultimately decided to score forecasts mainly based on the MSE, noting that the application of TADDA
\begin{quote}
``[...] is somewhat problematic. In particular, with the parameterization used for the evaluation, in the TADDA score, no model can outperform the no-change model predicting “no change in violence”. This means that a model with predictions very close to a constant no-change model would be preferred if evaluated according to this score, which the SC did not consider to be a good choice.'' \cite[p.~889]{Vesco2022}
\end{quote}
Indeed, as displayed, e.g., in Figure 2 of \cite{Vesco2022}, a \textit{no-change} model systematically predicting $\hat{y}_{t, s} = 0$ achieved the best average TADDA scores of all considered models. This led to the empirical conclusion that  ``the TADDA score appears to overly favor the simple no-change model'' \cite[892]{Vesco2022}. We provide a statistical explanation why a no-change forecast will often achieve better TADDA scores than more sophisticated forecasts, especially if these are designed to minimize the mean squared error.

The remainder of the paper is structured as follows. In Section \ref{Sec:Definitions} we provide the definition of the TADDA score. In Section \ref{Sec:Theory_General} we describe the statistical concept of the \textit{optimal point forecast} which enables us to study the incentives created by different scoring functions. The incentives implied by the TADDA score are discussed in Section \ref{Sec:Theory_TADDA}. In Section \ref{Sec:empirical_example} we provide an empirical illustration of our argument before concluding in Section \ref{Sec:discussion} with a discussion of alternative scoring concepts.

\section{Definition of the TADDA score}\label{Sec:Definitions}

\begin{figure}[htb]
    \begin{center}
            \includegraphics[width = \textwidth]{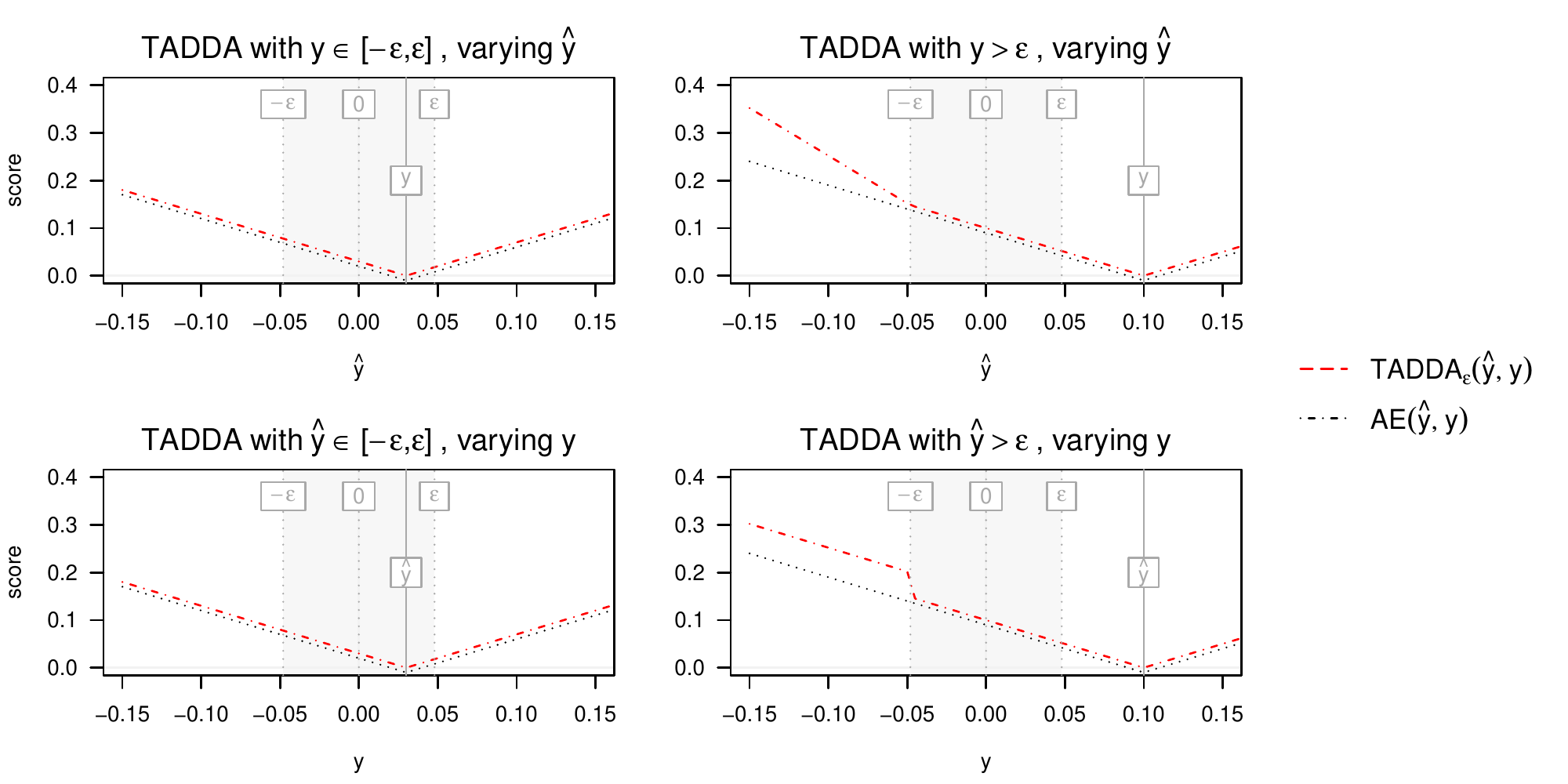}
    \end{center}
    \caption{Illustration of the absolute error (AE) and $\text{TADDA}_\epsilon$ with $\epsilon = 0.048$. Top row: as a function of $\hat{y}$ with fixed observation $y$. Bottom row: as a function of $y$ with fixed prediction $\hat{y}$. In each row we show one example where $\hat{y}$ or $y$, respectively are inside $[-\epsilon, \epsilon]$ and one where it is outside. The lines for the two scores are slightly shifted to avoid overplotting.}
    \label{fig:curves_L1}
\end{figure}

We start by providing the definition of the TADDA score, slightly adapting notation to our purposes. Denoting the outcome of interest by $y$  and the respective point forecast by $\hat{y}$, it is given by
\begin{equation} \label{eq:def_tadda}
    \text{TADDA}_\epsilon(\hat{y}, y) = |\hat{y} - y| \ \ + \ \ a_\epsilon(\hat{y}, y).
\end{equation}
Here, $a_\epsilon(\hat{y}, y)$ with $\epsilon > 0$ is a term reflecting \textit{direction augmentation} and defined as
\begin{equation}\label{eq:def_tadda1_L1_penalty}
a_\epsilon(\hat{y}, y) = 
 \begin{cases}
     \hat{y} - \epsilon & \text{ if } \ \ \hat{y} > \epsilon \text{ and } y < - \epsilon \\
     - \hat{y}  -\epsilon & \text{ if } \ \ \hat{y} < - \epsilon \text{ and } y > \epsilon \\
     0 & \text{ otherwise}.
\end{cases}
\end{equation}
The intuition is that a penalty is applied whenever $\hat{y}$ and $y$ are on opposite sides of a tolerance region $[-\epsilon, \epsilon]$, with $\epsilon$ chosen by the analyst. \cite{Vesco2022} use $\epsilon = 0.048$, which corresponds to a relative change of 5\% in observed fatalities. An illustration of the score with $\epsilon = 0.048$ is provided in Figure \ref{fig:curves_L1}. We added the absolute error (AE) for comparison, which corresponds to the absence of a penalty term $a_\epsilon(\hat{y}, y)$ or a very large value of $\epsilon$.

We note that the above specification of the score is denoted by TADDA1 in \cite{Vesco2022} and only one of several possible versions. Specifically, the score can also be based on a quadratic rather than linear distance, and an alternative handling of the tolerance region is implemented in a score called TADDA2. As it takes the most prominent role in \cite{Vesco2022} and as our findings largely translate to the other versions, we here restrict our attention to the score from equations \eqref{eq:def_tadda} and \eqref{eq:def_tadda1_L1_penalty}. A more general analysis covering different variations of TADDA is deferred to Supplement \ref{sec:proofs}.

\section{Scoring functions and incentives}\label{Sec:Theory_General}

Before we analyze the properties of the TADDA score, we introduce some helpful theoretical notions. Following \cite{Gneiting2011}, we conceive the prediction task as a decision problem under uncertainty. The forecaster issues a prediction $\hat{y}$ for a random variable $Y$, with the realized value $y$ yet unknown. The assessment of forecast quality is based on a \textit{scoring function} $s$, which returns a real number based on $\hat{y}$ and $y$. We orient it such that lower values correspond to better forecasts, meaning that the forecaster chooses her prediction $\hat{y}$ with the aim of minimizing her score. While the realized value $y$ of $Y$ is unknown at the time of prediction, we assume that the forecaster can describe her uncertainty about the future via a probability distribution $F$ for $Y$. When asked for a point prediction, the forecaster then has an incentive to issue the value $\hat{y}_{\text{OPF}}$ which under her predictive distribution $F$ yields the lowest expected score, i.e.,
$$
\hat{y}_\text{OPF} = \text{argmin}_{\hat{y} \in \mathbb{R}} \mathbb{E}_F[s(\hat{y}, Y)].
$$
In the statistical literature this choice of point forecast is often called the \textit{Bayes act} \citep{Gneiting2011}, but we will simply refer to it as the \textit{optimal point forecast} (OPF). Whenever a certain characteristic, or \textit{functional}, of the predictive distribution $F$ is the OPF of a specific scoring function, we say that the functional is \textit{elicited} by the scoring function. Vice versa, whenever a functional is the OPF under a given scoring function, the scoring function is said to be \textit{consistent} for this functional.

In two well-known cases, the OPF corresponds to measures of central tendency of the predictive distribution $F$. For the squared error (SE) $s(\hat y, y) = (y - \hat y)^2$, the OPF is given by the mean of the distribution $F$. For the absolute error (AE) $s(\hat y, y) = |y-\hat y|$, the OPF corresponds to the median of $F$. Hence, in the case of a skewed predictive distribution $F$, the squared and absolute errors usually imply different optimal point forecasts.

There is thus a duality between the scoring function and the functional of the forecaster’s predictive distribution which shall be elicited. For instance, a forecaster has an incentive to report a predictive mean when evaluated with the squared error. Conversely, for an evaluator it makes sense to apply the squared error when she knows that predictive means have been reported. By contrast, applying the absolute error for evaluation is incoherent in this situation: Had the forecaster known that the absolute error was applied, she would have reported the median rather than the mean of her predictive distribution. As pointed out by \cite{Gneiting2011} and \cite{Kolassa2020},  the definition of a forecasting task should therefore state either (i) one well-chosen scoring function which forecasters should aim to optimize or (ii) the requested functional of forecasters' predictive distributions, which can then be evaluated using one or several consistent scoring functions.

\subsection*{Simulation example}
\begin{figure}[htb]
    \centering
    \includegraphics[scale = 0.75]
    {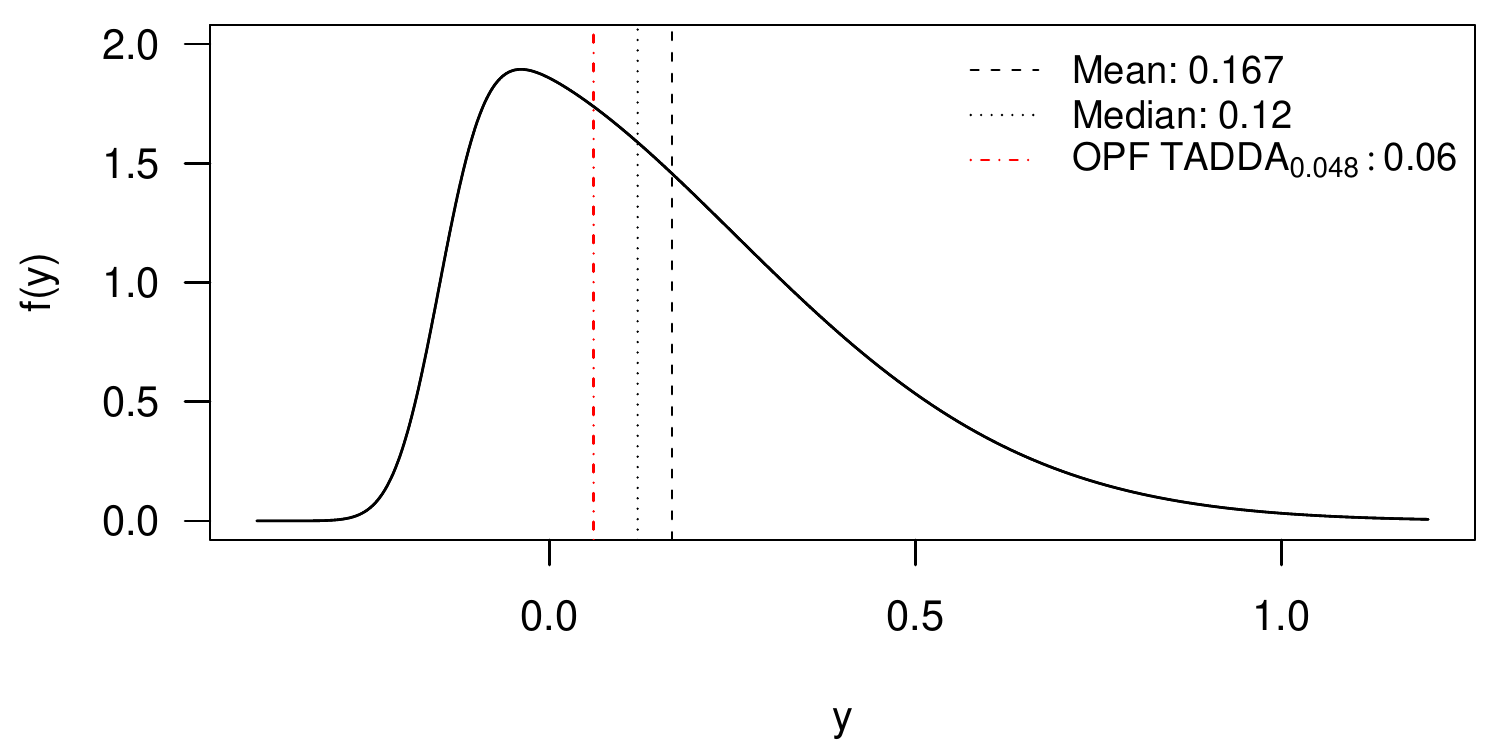}
    \caption{Probability density function of a skew normal distribution with parameters $\xi = -0.15$, $\omega = 0.4$ and $\alpha = 8$. The mean and median are marked by black dashed and dotted lines, respectively. The red lines show the optimal point forecast under $\text{TADDA}_{0.048}$, see Section \ref{Sec:Theory_TADDA}.}
    \label{fig:illustration}
\end{figure}

As an illustrative example we set $F$ to a skew normal distribution \citep[Chapter 2]{Azzalini2013} with parameters $\xi = -0.15$, $\omega = 0.4$ and $\alpha = 8$, see Figure \ref{fig:illustration} for the probability density function. As the distribution is skewed, its mean $\mu$ and median $m$ differ. They are given by $\mu = 0.167$ and $m = 0.120$, respectively. Via simulation it can be shown that under $F$, the expected absolute error when reporting the mean is $0.195$. When reporting the median instead, it can be lowered to $0.192$. Conversely, the expected squared error under $F$ is $0.060$ when reporting the mean, but $0.062$ for the median. As implied by theory, the predictive median is thus the better choice under the absolute error, and the predictive mean under the squared error (even though the magnitude of the differences may not be impressive).

\section{Incentives created by TADDA}\label{Sec:Theory_TADDA}
A natural question is now what the optimal point forecast under the TADDA score is, i.e., which functional of a forecaster's predictive distribution $F$ it elicits. 
The OPF for $\epsilon > 0$ is given by
\begin{equation}
    \hat{y}_{\text{OPF}} = \begin{cases}
    F^{-1}\{0.5 \times (1 + \pi_+)\} & \text{ if } \pi_- > 0.5 \times (1 + \pi_+) \ \ \ \ \ \ \ \ \ \ \ \ \, \text{\footnotesize high confidence that $Y < -\epsilon$} \\
     -\epsilon & \text{ if } 0.5 < \pi_- \le 0.5 \times (1 + \pi_+) \ \ \ \text{\footnotesize low confidence that $Y < -\epsilon$} \\
    F^{-1}(0.5) \ \ = \ \ m & \text{ if } \pi_- \leq 0.5 \text{ and } \pi_+ \leq 0.5  \ \ \ \ \ \ \ \ \ \, \text{\footnotesize unsure about sign of $Y$}\\
    \epsilon & \text{ if } 0.5 < \pi_+ \leq 0.5 \times (1 + \pi_-)  \ \ \ \text{\footnotesize low confidence that $Y > \epsilon$}\\
    F^{-1}\{0.5 \times (1 - \pi_-)\}& \text{ if } \pi_+ > 0.5 \times (1 + \pi_-).  \ \ \ \ \ \ \ \ \ \ \ \, \text{\footnotesize high confidence that $Y > \epsilon$} 
    \end{cases}
\label{eq:BA_TADDA1_L1}
\end{equation}
Here, we denote by $\pi_-$ and $\pi_+$ the predictive probability that $Y$ is below $-\epsilon$ and above $\epsilon$, respectively, and $F^{-1}$ is the predictive quantile function. The optimal point forecast thus depends on how confident the forecaster is about the sign of $Y$. If she is unsure with both $\pi_-$ and $\pi_+$ at most 0.5, the optimal point forecast is simply the predictive median $m$, which in this case is from $[-\epsilon, \epsilon]$. If she is more than 50\% confident that $Y > \epsilon$, i.e. $\pi_+ > 0.5$, but no more than a threshold of $0.5 \times (1 + \pi_-)$, the optimal point forecast equals the tolerance value $\epsilon$. And only if this threshold, which is somewhere between 0.5 and 2/3 depending on $\pi_-$, is exceeded, the optimal point forecast is larger than $\epsilon$. The forecaster then accepts the risk of a penalty from direction augmentation. However, the optimal point forecast is usually shifted towards $\epsilon$  relative to $m$ as it consists of the $0.5 \times (1 - \pi_-)$ rather than the 0.5 quantile of $F$. All of this reflects that the forecaster tries to avoid strong penalties from the direction augmentation term. For the case $\pi_- > 0.5$, the previous description translates via symmetry. Proofs can be found in Appendix \ref{sec:proofs}.

\subsection*{Simulation example \textit{(continued)}}
We return to the example from the previous section to illustrate our results. For $\text{TADDA}_{0.048}$, the optimal point forecast is approximately 0.06. The forecaster is thus sufficiently confident that $Y > 0.048$ to issue a forecast outside of the tolerance region (specifically, she is 62\% sure that $Y > \epsilon$). Compared to the predictive mean and median, however, she stays quite close to $\epsilon$. To contrast TADDA and the commonly used absolute and squared errors, Table \ref{tab:scores} summarizes the respective optimal point forecasts (Panel a) and their expected scores under the different metrics (Panel b). We moreover added a forecast of zero, corresponding to the no-change model from \cite{Vesco2022}. As implied by theory, the lowest expectation for each score is achieved by the respective optimal point forecast. A central observation is that the zero forecast is a better choice than the predictive mean under TADDA. As most submitted forecasting models were optimized with respect to MSE \cite[p.~529]{Hegre2022}, this may explain why none of them was able to outperform the no-change forecast on TADDA.

\begin{table}[htb]
\centering
\footnotesize
\caption{Expected scores for different combinations of reported functional and scoring function under a skew normal distribution $F$ with $\xi = -0.15$, $\omega = 0.4$ and $\alpha = 8$ (see Figure \ref{fig:illustration}). The best (lowest) expected score for each scoring rule is highlighted in bold.}
\begin{tabular}{lccc}
    \multicolumn{4}{c}{(a) Optimal point forecasts under different scoring rules} \\[2pt]
  \toprule
  & AE & SE & $\text{TADDA}_{0.048}$ \\
  \midrule
  Functional & median & mean & see equation \eqref{eq:BA_TADDA1_L1} \\
  Value &  0.120 & 0.167 & 0.060 \\ \bottomrule
  \\
    \multicolumn{4}{c}{(b) Expected scores for different functionals of the distribution $F$}\\ [2pt]
  \toprule
  Functional & AE & SE & $\text{TADDA}_{0.048}$ \\ 
  \midrule
  Median & \textbf{0.192} & 0.062 & 0.207 \\ 
  Mean & 0.195 & \textbf{0.060} & 0.219 \\ 
  OPF $\text{TADDA}_{0.048}$ & 0.198 & 0.071 & \textbf{0.200} \\ 
  Zero & 0.216 & 0.088 & 0.216 \\ 
  \bottomrule
\end{tabular}
\label{tab:scores}
\end{table}

\section{Empirical example}\label{Sec:empirical_example}

In the previous section, we demonstrated that different scoring functions generally lead to different optimal point forecasts. In particular, the predictive mean is usually a sub-optimal choice under $\text{TADDA}$. Even if a forecaster's predictive distribution $F$ equals the true distribution, its mean $\mu$ may in expectation yield worse TADDA scores than a zero forecast. We now illustrate that this is not merely a theoretical peculiarity, but can translate to applied settings like the ViEWS competition. We here limit ourselves to the comparison of TADDA and SE, i.e., the measures used in \cite{Vesco2022}.

The ViEWS competition comprised two prediction tasks. Task 1 consisted of monthly real-time forecasts for the period October 2020 through March 2021, the true future at the time of the deadline \citep{Hegre2022}. Given the inherent sparsity of armed conflict data, this was complemented with a more comprehensive set of retrospectively generated forecasts. Task 2 therefore consisted of a 36-month period (January 2017 through December 2019), for which teams generated forecasts at lead times of $s = 2, \dots, 7$ corresponding to the lead times of task 1. In terms of $\text{TADDA}_{0.048}$, none of the contributed models achieved an improvement over the no-change forecast for either of the two tasks (compare Footnote 3 in \citealt{Vesco2022}). In the following, we focus on the more extensive task 2 and revisit the evaluation at the country-month level presented in Table 2 of \cite{Hegre2022}.

We generate a simple probabilistic forecast based on recent observations, from which we derive two sets of point forecasts optimized for SE and TADDA, respectively. Recall that, for a lead time of $s$ months, under the no-change forecast we predict that the number of fatalities $X_t$ in month $t$ are identical to the last known value $x_{t - s}$. This implies a prediction of $\hat{y}_{t, s} = 0$ for the log-change. For our probabilistic forecast, we extend this to the last $w$ observations $\{x_{t - s - i},\; i = 0, \dots, w - 1\}$ and compute the log-changes they would imply relative to the last value $x_{t - s}$. We denote these by 
$$
y^*_{t, s, i} = \log(x_{t - s} + 1) - \log(x_{t - s - i} + 1), \ \ \ i = 0, \dots, w - 1.
$$
We then predict that $Y_{t, s}$ takes on each of the values $y^*_{t, s, i}, i = 1, \dots, w$ with equal probability $1/w$. 
From this distribution, we obtain the optimal point forecasts with respect to the SE and $\text{TADDA}_{0.048}$, i.e., the predictive means and the functionals from equation \eqref{eq:BA_TADDA1_L1}. To evaluate quantiles we use the inverse of the empirical cumulative distribution  function as implemented in the function \texttt{quantile(..., type = 1)} in R \citep{R2021}.

\begin{figure}
    \centering
    \includegraphics[width = 0.95\textwidth]{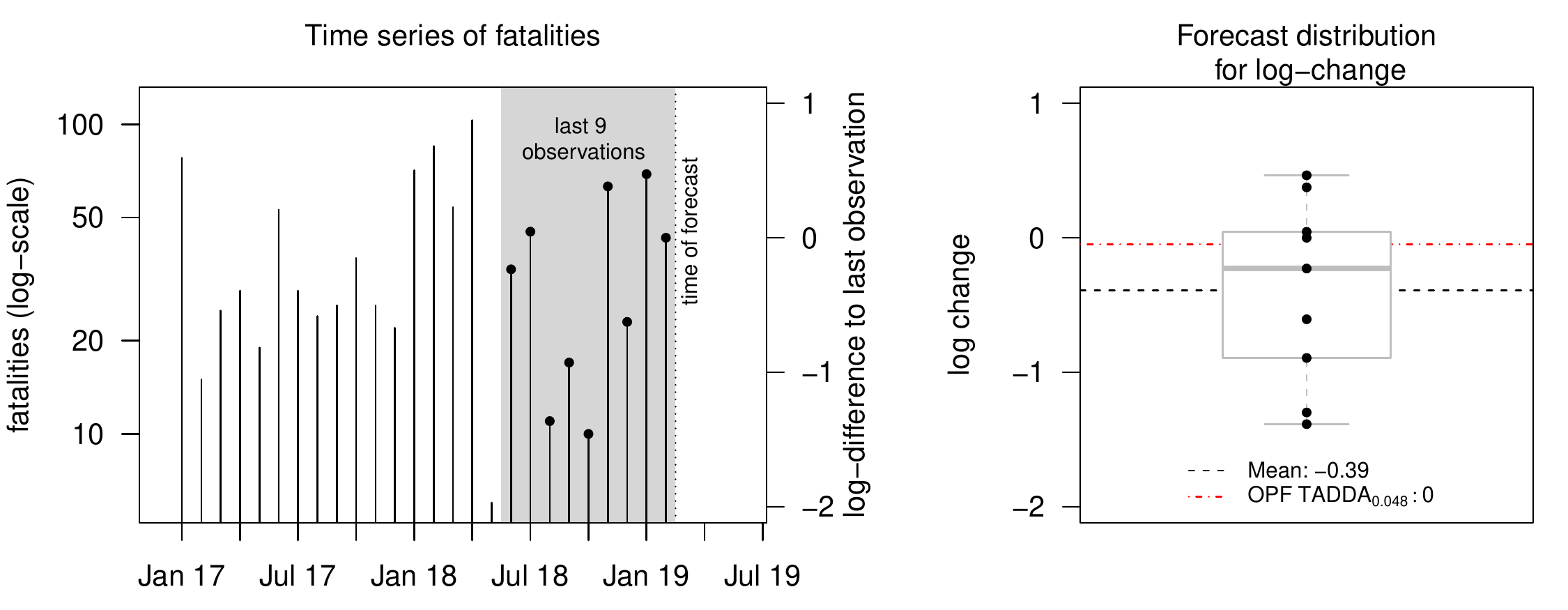}
    \caption{Construction of the forecast distribution at time $t - s$ (here: February 2019): The last $w = 9$ available monthly numbers of fatalities $x_{t - s - i}$, $i = 0, \dots, 8$, are extracted (left panel) and for each of them the log-change $y^*_{t, s, i}$ relative to the latest observation $x_{t - s}$ is computed (note that one of them is 0 by construction). The resulting values are used to define a nine-point predictive distribution for $Y_{t, s}$ (right panel). The example data used for the illustration are from Mali.}
    \label{fig:example_mali}
\end{figure}

Based on the calibration period (January 2014 through December 2016), the optimal window size in terms of $\text{TADDA}_{0.048}$ is $w = 5$ months. Indeed, this choice also yields the best results for the evaluation period. However, as this corresponds to very coarse predictive distributions, we here use $w = 9$ for illustration. The $y^*_{t, s, i}$ can then be seen as the 0\%, 12.5\%, 25\%, \dots, 87.5\%, 100\% quantiles of the predictive distribution. Figure \ref{fig:example_mali} illustrates how the predictive distribution for $Y_{t, s}$ arises from the $w = 9$ observations leading up to the time of prediction $t - s$.

\begin{table}[htbp]
  \centering
  \caption{Average evaluation scores for the different lead times $s$ and window length $w=9$ in the prediction period January 2017 to December 2019. The table includes scores for the predictive means and optimal point forecasts for $\text{TADDA}_{0.048}$ from our model. For comparison we also included the no-change model and the ViEWS ensemble (Table 2 in \citealt{Vesco2022}). The bottom line contains the column means.}
    \footnotesize
    \begin{tabular}{ccccccccc}
    \toprule
    & \multicolumn{4}{c}{MSE} & \multicolumn{4}{c}{TADDA$_{0.048}$} \\\cmidrule[0.5pt](lr{.75em}){2-5} \cmidrule[0.5pt](lr{.75em}){6-9}
    $s$ & Pred. mean & OPF TADDA$_{0.048}$ & ViEWS & No-change & Pred. mean & OPF TADDA$_{0.048}$ & ViEWS & No-change \\ \midrule
    2    & 0.487 & 0.561 & 0.504 & 0.674 & 0.357 & 0.335 & 0.371 & 0.340 \\
    3    & 0.525 & 0.619 & 0.551 & 0.773 & 0.368 & 0.350 & 0.379 & 0.365 \\
    4    & 0.556 & 0.651 & 0.579 & 0.841 & 0.372 & 0.359 & 0.394 & 0.391 \\
    5    & 0.587 & 0.673 & 0.548 & 0.807 & 0.388 & 0.369 & 0.381 & 0.375 \\
    6    & 0.629 & 0.720 & 0.573 & 0.841 & 0.399 & 0.380 & 0.386 & 0.384 \\
    7    & 0.664 & 0.749 & 0.599 & 0.864 & 0.410 & 0.388 & 0.400 & 0.389 \\\midrule
     & 0.575 & 0.662 & 0.559 & 0.800 & 0.382 & 0.363 & 0.385 & 0.374 \\\bottomrule
    \end{tabular}%
  \label{tab:eval_pred_w9}%
\end{table}%

Table \ref{tab:eval_pred_w9} summarizes the performance of both types of point forecasts derived from our model, along with the respective scores of the ViEWS ensemble and the no-change model from \cite{Vesco2022}. The ViEWS ensemble leads the field in terms of MSE, in particular for longer lead times. The predictive means from our model outperform the respective TADDA$_{0.048}$ OPFs, while the no-change baseline shows the weakest performance. Regarding TADDA$_{0.048}$, the picture is quite different: while the optimal point forecasts from our model outperform the no-change model consistently across all horizons, the predictive means and the ViEWS ensemble have worse average scores. These results demonstrate that each forecasting approach performs well under the score it is optimized for. The ViEWS ensemble arose from models which target MSE, and its weights were likewise determined based on MSE \citep{Hegre2022}. Our simple model is almost competitive with the ensemble when optimized for MSE by using predictive means. The TADDA$_{0.048}$ OPFs, on the other hand,  yield only mediocre performance in terms of MSE. Yet, they are the only ones to outperform the no-change model under TADDA$_{0.048}$, i.e., the score for which they were optimized.

The fact that the $\text{TADDA}_{0.048}$ OPFs from our model outperform the predictive means under $\text{TADDA}_{0.048}$, while the opposite is the case under the SE, is robust to the choice of window length $w > 1$. The $\text{TADDA}_{0.048}$ OPFs also outperform the no-change model for any window length $w = 2, \dots, 8$. For $w = 3, \dots, 7$, the predictive means from our model also achieve slight improvements over the no-change model, indicating that it is not impossible to outperform the no-change model under TADDA while optimizing for MSE.

To improve our intuition for the behavior of the compared forecasting approaches, we conclude by contrasting some of their characteristics. Table \ref{tab:emp_distr_mean_BA_TADDA} describes the empirical distributions of the point forecasts and observed log-changes. The most notable pattern is that a majority of observed and predicted log-changes are zero. More precisely, this is the case for 72.0\% of the observations, which is one reason why the no-change forecast is not straightforward to beat. Concerning the forecasts, 
78.8\% of the TADDA$_{0.048}$ OPFs and 63.9\% of the predictive means from our model are zero. The TADDA$_{0.048}$ OPFs are much less frequently outside the interval $[-0.048, 0.048]$ than the predictive means. This is beneficial for their TADDA scores, as they avoid the risk of penalties from direction augmentation. The no-change forecasts, too, benefit from this characteristic of the score.

\begin{table}[htbp]
  \centering
  \caption{Empirical quantiles of the predicitve means and TADDA$_{0.048}$ optimal point forecasts as well as the true log-changes across all lead times $s=2,...,7$ in the prediction period January 2017 to December 2019.}
  \footnotesize
    \begin{tabular}{lcccccccccc} \toprule
          & 5\%   & 10\% & 15\% & 20\%  & $\cdots$ & 75\%  & 80\% & 85\% & 90\% & 95\% \\\midrule
    Mean   & --1.038 & --0.510 & --0.068 & 0 & $\cdots$ & 0 & 0.033 & 0.256 & 0.462 & 0.877 \\
    OPF TADDA$_{0.048}$ & --1.099 & --0.175 & 0 & 0 & $\cdots$ & 0 & 0 & 0 & 0 & 0.462  \\
    True log-changes & --1.386 & --0.588 & 0 & 0 & $\cdots$ & 0 & 0 & 0 & 0.693 & 1.447  \\\bottomrule
    \end{tabular}%
  \label{tab:emp_distr_mean_BA_TADDA}%
\end{table}

\section{Discussion}\label{Sec:discussion}
The theoretical and empirical results from the previous sections underscore the duality between scoring functions and optimal point forecasts. This has two main implications. Firstly, whenever a scoring rule is chosen, it should be assessed whether the functional of a forecaster's predictive distribution it elicits is of interest. As an example, expected (mean) costs may be relevant in financial contexts, and can be elicited via the squared error. Concerning the TADDA score, it is unclear whether the hard-to-interpret functional provided in equation \eqref{eq:BA_TADDA1_L1} is of practical relevance. Given that the resulting point forecasts in absolute value rarely exceed the tolerance value $\epsilon$, they may be considered overly conservative. This is echoed in the judgment of the independent scoring committee cited in Section \ref{sec:introduction}. Secondly, while applying different evaluation scores in parallel can yield a more comprehensive picture of performance, it should be ensured that they all elicit the same functional (see also \citealt{Gneiting2011} and \citealt{Kolassa2020}). Otherwise, conflicting incentives arise, and forecasters lack a clear objective.

A natural question is whether other scores can be conceived which reward point forecasts for having the correct sign, but elicit standard and interpretable functionals of the forecasters' predictive distributions. General construction principles for scoring functions to elicit predictive means have been established in \cite{Ehm2016}. It seems feasible to construct variations of TADDA which elicit predictive means or medians, but a detailed discussion is outside the scope of this note and will be provided elsewhere. Another possibility is to evaluate a model's point forecasts via the squared or absolute error, and complement this with an assessment of the model's predicted probability of a positive outcome. The latter could be assessed using the Brier score, a widely employed scoring rule for binary targets.

Finally, a more general alternative to the evaluation of point forecasts as performed in the \textit{VIEWS} challenge is to collect and score probabilistic forecasts. This way, no choice concerning an appropriate functional to summarize predictive distributions would be necessary. The potential of a probabilistic approach has already been evoked in the outlook of \cite{Vesco2022}, and \cite{Brandt2022} provide some results based on the continuous ranked probability score (CRPS). There is a rich body of literature on methods for probabilistic forecast evaluation \citep{Gneiting2014}, and these are widely used in challenge-based formats, e.g., in  meteorological \citep{Vitart2022}, epidemiological \citep{Bracher2021} and energy forecasting \citep{Hong2016}.

\section*{Data and code availability statement} Code and data to reproduce the presented results can be found at \url{https://github.com/KITmetricslab/tadda}.

\section*{Funding details} Johannes Bracher was supported by the Helmholtz Information and Data Science Project \textit{SIMCARD} as well as Deutsche Forschungsgemeinschaft (DFG, German Research Foundation) -- project number 512483310. Sebastian Lerch gratefully acknowledges support by the Vector Stiftung through the Young Investigator Group ``Artificial Intelligence for Probabilistic Weather Forecasting''. Melanie Schienle gratefully acknowledges funding by the HEiKA EXC-scouting initiative.



\begin{thebibliography}{}

\bibitem[Azzalini, 2013]{Azzalini2013}
Azzalini, A. (2013).
\newblock {\em The Skew-Normal and Related Families}.
\newblock Institute of Mathematical Statistics Monographs. Cambridge University
  Press.

\bibitem[Bracher et~al., 2021]{Bracher2021}
Bracher, J., Ray, E.~L., Gneiting, T., and Reich, N.~G. (2021).
\newblock Evaluating epidemic forecasts in an interval format.
\newblock {\em PLOS Computational Biology}, 17(2):e1008618.

\bibitem[Brandt et~al., 2022]{Brandt2022}
Brandt, P.~T., D’Orazio, V., Khan, L., Li, Y.-F., Osorio, J., and Sianan, M.
  (2022).
\newblock Conflict forecasting with event data and spatio-temporal graph
  convolutional networks.
\newblock {\em International Interactions}, 48(4):800--822.

\bibitem[Ehm et~al., 2016]{Ehm2016}
Ehm, W., Gneiting, T., Jordan, A., and Krüger, F. (2016).
\newblock Of quantiles and expectiles: consistent scoring functions, choquet
  representations and forecast rankings.
\newblock {\em Journal of the Royal Statistical Society. Series B (Statistical
  Methodology)}, 78(3):505--562.

\bibitem[Gneiting, 2011]{Gneiting2011}
Gneiting, T. (2011).
\newblock Making and evaluating point forecasts.
\newblock {\em Journal of the American Statistical Association},
  (494):746--762.

\bibitem[Gneiting and Katzfuss, 2014]{Gneiting2014}
Gneiting, T. and Katzfuss, M. (2014).
\newblock Probabilistic forecasting.
\newblock {\em Annual Review of Statistics and Its Application}, 1(1):125--151.

\bibitem[Hegre et~al., 2019]{Hegre2019}
Hegre, H., Allansson, M., Basedau, M., Colaresi, M., Croicu, M., Fjelde, H.,
  Hoyles, F., Hultman, L., H{\"o}gbladh, S., Jansen, R., et~al. (2019).
\newblock Views: A political violence early-warning system.
\newblock {\em Journal of peace research}, 56(2):155--174.

\bibitem[Hegre et~al., 2022]{Hegre2022}
Hegre, H., Vesco, P., and Colaresi, M. (2022).
\newblock Lessons from an escalation prediction competition.
\newblock {\em International Interactions}, 48(4):521--554.

\bibitem[Hong et~al., 2016]{Hong2016}
Hong, T., Pinson, P., Fan, S., Zareipour, H., Troccoli, A., and Hyndman, R.~J.
  (2016).
\newblock Probabilistic energy forecasting: Global energy forecasting
  competition 2014 and beyond.
\newblock {\em International Journal of Forecasting}, 32(3):896--913.

\bibitem[Kolassa, 2020]{Kolassa2020}
Kolassa, S. (2020).
\newblock Why the “best” point forecast depends on the error or accuracy
  measure.
\newblock {\em International Journal of Forecasting}, 36(1):208--211.
\newblock M4 Competition.

\bibitem[{R Core Team}, 2021]{R2021}
{R Core Team} (2021).
\newblock {\em R: A Language and Environment for Statistical Computing}.
\newblock R Foundation for Statistical Computing, Vienna, Austria.

\bibitem[Vesco et~al., 2022]{Vesco2022}
Vesco, P., Hegre, H., Colaresi, M., Jansen, R.~B., Lo, A., Reisch, G., and
  Weidmann, N.~B. (2022).
\newblock United they stand: Findings from an escalation prediction
  competition.
\newblock {\em International Interactions}, 48(4):860--896.

\bibitem[Vitart et~al., 2022]{Vitart2022}
Vitart, F., Robertson, A.~W., Spring, A., Pinault, F., Roškar, R., Cao, W.,
  Bech, S., Bienkowski, A., Caltabiano, N., De~Coning, E., Denis, B., Dirkson,
  A., Dramsch, J., Dueben, P., Gierschendorf, J., Kim, H.~S., Nowak, K.,
  Landry, D., Lledó, L., Palma, L., Rasp, S., and Zhou, S. (2022).
\newblock Outcomes of the {WMO} {Prize} {Challenge} to {Improve} {Subseasonal}
  to {Seasonal} {Predictions} {Using} {Artificial} {Intelligence}.
\newblock {\em Bulletin of the American Meteorological Society},
  103(12):E2878--E2886.

\end{thebibliography}

\newpage

\renewcommand{\thepage}{S\arabic{page}}
\renewcommand{\thesection}{S\arabic{section}}
\renewcommand{\thetable}{S\arabic{table}}
\renewcommand{\thefigure}{S\arabic{figure}}
\renewcommand{\theequation}{S\arabic{equation}}

\newpage

\appendix

\section{Proofs and results for other variants of TADDA}\label{sec:proofs}

In this appendix, we provide proofs for the results on the optimal point forecasts for TADDA from the main manuscript as well as corresponding results for two further variations of the score. To distinguish them clearly, we will denote the variant discussed in the main text by $\text{TADDA1}_\epsilon^{\text{L1}}$. We will further address the same score based on an L2 rather than L1 distance, denoted by $\text{TADDA1}_\epsilon^{\text{L2}}$. Lastly, we consider an alternative handling of the tolerance region proposed by \cite{Vesco2022} under L1 distance. Following the convention from \cite{Vesco2022}, we denote the latter by $\text{TADDA2}_\epsilon^{\text{L1}}$. As mentioned in Section \ref{Sec:Theory_TADDA}, we will use $\pi_-:=\text{Pr}_F(Y < -\epsilon)$ and $\pi_+:=\text{Pr}_F(Y > \epsilon)$.

\subsection{Derivation of the optimal point forecast for $\text{TADDA1}_\epsilon^{\text{L1}}$}

\textit{Result:} The optimal point forecast for
$$
\text{TADDA1}_\epsilon^{\text{L1}}(\hat{y}, y) = |\hat{y} - y| + a_\epsilon(\hat{y}, y),
$$
where $a_\epsilon(\hat{y}, y)$ is defined as in equation \eqref{eq:def_tadda1_L1_penalty} is given by
$$
\hat{y}_\text{OPF} = \begin{cases}
    F^{-1}\{0.5 \times (1 + \pi_+)\} & \text{ if } \pi_- \ge 0.5 \times (1 + \pi_+)\\
     -\epsilon & \text{ if } 0.5 < \pi_- < 0.5 \times (1 + \pi_+)\\
    F^{-1}(0.5) \ \ \ = m & \text{ if } \pi_- \leq 0.5 \text{ and } \pi_+ \leq 0.5\\
    \epsilon & \text{ if } 0.5 < \pi_+ \leq 0.5 \times (1 + \pi_-)\\
    F^{-1}\{0.5 \times (1 - \pi_-)\}& \text{ if } \pi_+ > 0.5 \times (1 + \pi_-).
    \end{cases}
$$
Note that the definition of the direction augmentation term is equivalent to
\begin{equation}
a_\epsilon(\hat{y}, y) = 
 \begin{cases}
     |\hat{y} - \epsilon| & \text{ if } \ \ \hat{y} > \epsilon \text{ and } y < - \epsilon \\
     |\hat{y} + \epsilon| & \text{ if } \ \ \hat{y} < - \epsilon \text{ and } y > \epsilon \\
     0 & \text{ otherwise},
 \end{cases}
 \label{eq:reformulation_a}
\end{equation}
which we will use in the following.
\\

\textit{Proof:} The expectation of $\text{TADDA1}_\epsilon^{\text{L1}}(\hat{y}, Y)$ under the predictive distribution $F$ reads
$$
\mathbb{E}_F[\text{TADDA1}_\epsilon^{\text{L1}}(\hat{y}, Y)] = \underbrace{\mathbb{E}_F|\hat{y} - Y|}_{\text{(a)}} \ \ + \ \ \underbrace{\mathbb{E}_F[a_\epsilon(\hat{y}, Y)]}_{\text{(b)}}.
$$
Both terms (a) and (b) are obviously non-negative which translates directly to their expectation. Term (a) is minimized by $\hat{y} = m$ and increases monotonically to either side of $m$. Further, term (b) is zero 
for $\hat{y} \in [-\epsilon, \epsilon]$ and non-negative for $\hat{y} \notin [-\epsilon, \epsilon]$. Note that it is zero in particular for $\hat{y}=\pm \epsilon$. We show the result via proof by cases.

\textbf{Case 1 $(-\epsilon \leq m \leq \epsilon):$} Due to the above-mentioned characteristics of terms (a) and (b), the optimal point forecast is given by $\hat{y}_\text{OPF} = m$.

\begin{figure}
    \centering
    \includegraphics[width = 0.65\textwidth]{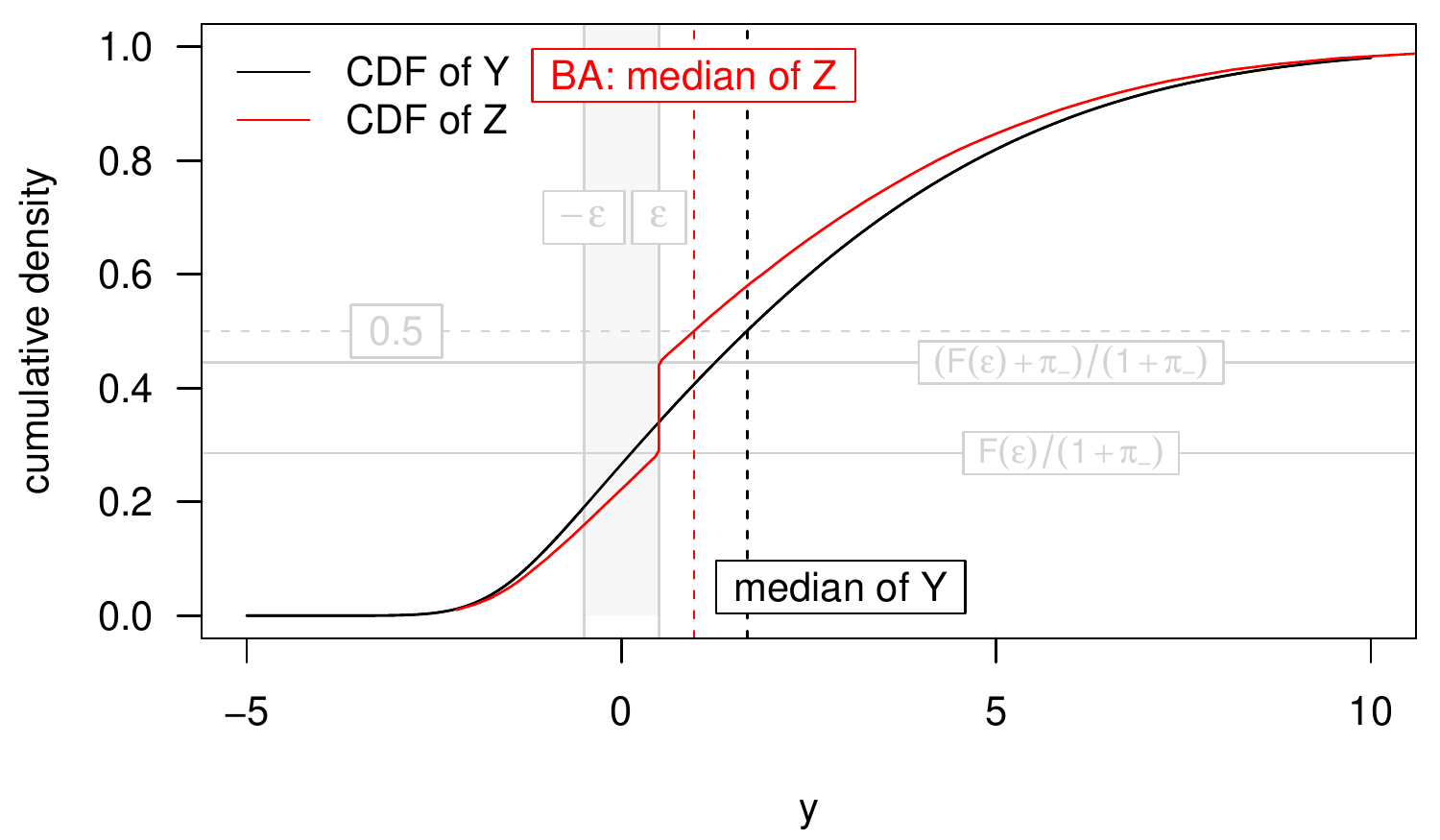}
    \caption{Illustration of the CDFs $F$ and $G$ for Case 2 of $\text{TADDA1}_\epsilon^{\text{L1}}$, where $\epsilon > 0$ and $ m > \epsilon$.}
    \label{fig:F_vs_G_epsilon_m>epsilon}
\end{figure}

\textbf{Case 2 $(m > \epsilon):$} For all $\hat{y} < \epsilon$, term (a) satisfies
$$
\mathbb{E}_F|\epsilon - Y| \leq \mathbb{E}_F|\hat{y} - Y|.
$$
Additionally, we know that no $\hat{y} < \epsilon$ can achieve a smaller value of term (b) than $\hat{y} = \epsilon$. Combined, we have that
$$
\mathbb{E}_F[\text{TADDA1}_\epsilon^{\text{L1}}(\epsilon, Y)] \leq \mathbb{E}_F[\text{TADDA1}_\epsilon^{\text{L1}}(\hat{y}, Y)] \ \text{ for all } \hat{y} < \epsilon.
$$
We can thus restrict our search for the optimal point forecast to $\hat{y} \geq \epsilon$. On this segment of the real line, the expected score is given by (see equation \eqref{eq:reformulation_a})
\begin{align*}
\mathbb{E}_F[\text{TADDA1}_\epsilon^{\text{L1}}(\hat{y}, Y)] & = \mathbb{E}_F|\hat{y} - Y| \ \ + \ \ \pi_- \times \mathbb{E}_F|\hat{y} - \epsilon|\\
& \propto \mathbb{E}_F|\hat{y} - Z|,
\end{align*}
where
$$
Z = \begin{cases}
Y & \text{ with probability } \frac{1}{1 + \pi_-}\\
\epsilon & \text{ with probability } \frac{\pi_-}{1 + \pi_-}.
\end{cases}
$$
The optimal choice for $\hat{y}$ is consequently the median $m_Z$ of $Z$.
The CDF of $Z$ is given by
\begin{equation}
G(z) = \begin{cases}
\frac{1}{1 + \pi_-} \times F(z) & \ \ \text{ for } \ \ z < \epsilon\\
\frac{1}{1 + \pi_-} \times F(z)+ \frac{\pi_-}{1 + \pi_-} & \ \ \text{ for } \ \ z \geq \epsilon,
\end{cases}\label{eq:F_Z_epsilon}
\end{equation}
see Figure \ref{fig:F_vs_G_epsilon_m>epsilon} for an illustration.
Since $m>\epsilon$ by assumption of Case 2, we know that $F(\epsilon)<0.5$. The median $m_Z$ of $Z$ is obtained by setting $G(m_Z) = 0.5$ in equation \eqref{eq:F_Z_epsilon} and solving for $m_Z$. This yields
\begin{equation*}
\hat{y}_\text{OPF} = m_Z = \begin{cases}
\epsilon & \text{ if } \frac{F(\epsilon) + \pi_-}{1 + \pi_-} \geq 0.5\\
F^{-1}\{0.5 \times (1 - \pi_-)\} & \text{ otherwise }.
\end{cases}
\end{equation*}
We can substitute $F(\epsilon) = 1 - \pi_+$ in the condition, which leads to
$$
\pi_+ \leq 0.5 \times(1 + \pi_-).
$$
In conjunction with $\pi_+ > 0.5$ (by assumption of Case 2), this gives the stated part of the result.

\begin{figure}
    \centering
    \includegraphics[width = 0.65\textwidth]{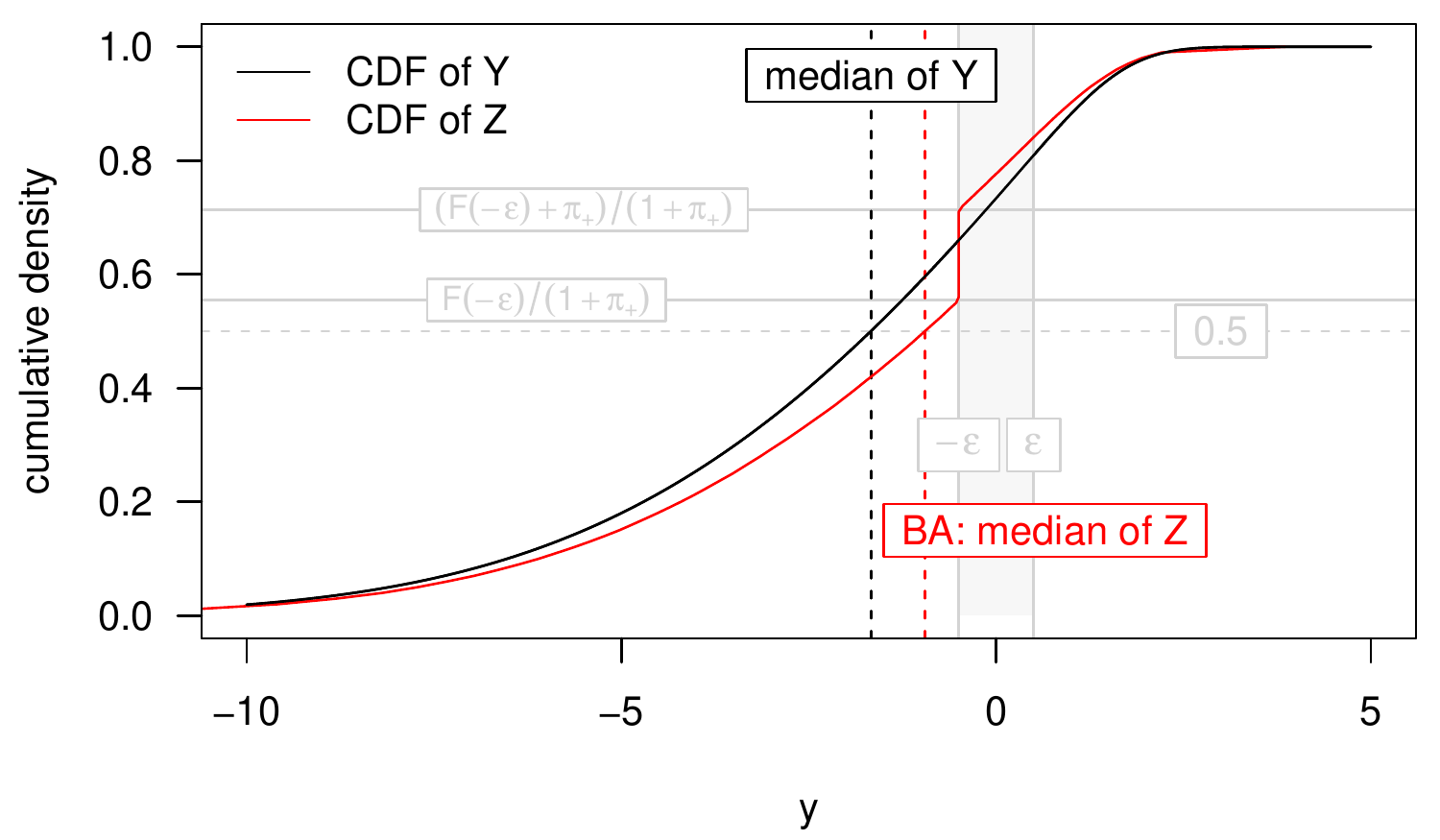}
    \caption{Illustration of the CDFs $F$ and $G$ for Case 3 of $\text{TADDA1}_\epsilon^{\text{L1}}$, where $\epsilon > 0$ and $ m < -\epsilon$.}
    \label{fig:F_vs_G_epsilon_m<-epsilon}
\end{figure}

\textbf{Case 3 $(m < -\epsilon):$} Analogously to Case 2, we have that
$$
\mathbb{E}_F[\text{TADDA1}_\epsilon^{\text{L1}}(-\epsilon, Y)] \leq \mathbb{E}_F[\text{TADDA1}_\epsilon^{\text{L1}}(\hat{y}, Y)] \ \text{ for all } \hat{y} > -\epsilon.
$$
Therefore, we can restrict our search for the optimal point forecast to $\hat{y} \leq - \epsilon$. For this segment of the real line, we obtain the expected score via
\begin{align*}
\mathbb{E}_F[\text{TADDA1}_\epsilon^{\text{L1}}(\hat{y}, Y)] & = \mathbb{E}_F|\hat{y} - Y| \ \ + \ \ \pi_+ \times \mathbb{E}_F|\hat{y} - (-\epsilon)|\\
& \propto \mathbb{E}_F|\hat{y} - Z|,
\end{align*}
where
$$
Z = \begin{cases}
Y & \text{ with probability } \frac{1}{1 + \pi_+}\\
-\epsilon & \text{ with probability } \frac{\pi_+}{1 + \pi_+}.
\end{cases}
$$
Again, the optimal choice for $\hat{y}$ is the median $m_Z$ of $Z$. The CDF of $Z$ is given by
\begin{equation}
G(z) = \begin{cases}
\frac{1}{1 + \pi_+} \times F(z)& \ \ \text{ for } \ \ z < -\epsilon\\
\frac{1}{1 + \pi_+} \times F(z) + \frac{\pi_+}{1 + \pi_+}& \ \ \text{ for } \ \ z \geq -\epsilon
\end{cases}\label{eq:F_Z_-epsilon}
\end{equation}
 and visualized in Figure \ref{fig:F_vs_G_epsilon_m<-epsilon}. Setting $G(m_Z) = 0.5$ in equation \eqref{eq:F_Z_-epsilon} and solving for $m_Z$ leads to 
$$
\hat{y}_\text{OPF} = m_Z = \begin{cases}
F^{-1}\{0.5 \times (1 + \pi_+)\} & \text{ if } \frac{\text{Pr}_F(Y<-\epsilon)}{1 + \pi_+} \ge 0.5 \\
-\epsilon & \text{ otherwise}.
\end{cases}
$$
We can rewrite the condition as 
$$
\frac{\pi_-}{1 + \pi_+} \ge 0.5 \;\; \Longleftrightarrow \;\; \pi_- \ge 0.5 \times (1 + \pi_+).
$$
This concludes the proof.

\subsection{Derivation of the optimal point forecast for $\text{TADDA1}^\text{L2}_\epsilon$}

\textit{Result:} The optimal point forecast for 
$$\text{TADDA1}^\text{L2}_\epsilon(\hat{y}, y) = (\hat{y}-y)^2 + a_\epsilon(\hat{y}, y),$$
where $a_\epsilon(\hat{y}, y)$ is defined as
\begin{equation*}\label{eq:def_TADDA1_L2_penalty}
a_\epsilon(\hat{y}, y) = \begin{cases}
    (\hat{y} - \epsilon)^2 & \text{ if } \ \ \hat{y} > \epsilon \text{ and }y < -\epsilon\\
    (\hat{y} + \epsilon)^2 & \text{ if } \ \ \hat{y} < - \epsilon \text{ and } y > \epsilon \\
    0 & \text{ otherwise},
\end{cases}
\end{equation*}
is given by
$$
\hat{y}_\text{OPF} =\begin{cases}
\mu \times \frac{1}{1 + \pi_+} - \epsilon \times \frac{\pi_+}{1 + \pi_+} & \text{ if } \mu < -\epsilon\\
\mu & \text{ if } -\epsilon \leq \mu \leq \epsilon\\
\mu \times \frac{1}{1 + \pi_-} + \epsilon \times \frac{\pi_-}{1 + \pi_-}& \text{ if } \mu > \epsilon.
\end{cases}
$$
\\

\textit{Proof:}
The expectation of $\text{TADDA1}^\text{L2}_\epsilon(\hat{y}, Y)$ reads
$$
\mathbb{E}_F[\text{TADDA1}^\text{L2}_\epsilon(\hat{y}, Y)] = \underbrace{\mathbb{E}_F[(\hat{y} - Y)^2]}_{\text{(a)}} \ \ + \ \ \underbrace{\mathbb{E}_F[a_\epsilon(\hat{y}, Y)]}_{\text{(b)}}.
$$
As in the L1 case, terms (a) and (b) are non-negative. Term (a) is minimized by the mean $\mu$ of $F$, i.e., $\hat{y} = \mu$, and increases monotonically to each side of $\mu$. Further, term (b) is zero for $\hat{y} \in [-\epsilon, \epsilon]$ and non-negative for $\hat{y} \notin [-\epsilon, \epsilon]$. Note that it is zero in particular for $\hat{y}=\pm \epsilon$. Again, we show the result via proof by cases.

\textbf{Case 1 $(-\epsilon \leq \mu \leq \epsilon):$} Due to the above-mentioned characteristics of terms (a) and (b), the optimal point forecast is given by $\hat{y}_\text{OPF} = \mu$.

\textbf{Case 2 $(\mu > \epsilon):$} For all $\hat{y} < \epsilon$, term (a) satisfies
$$
\mathbb{E}_F[(\epsilon - Y)^2] \leq \mathbb{E}_F[(\hat{y} - Y)^2],
$$
and we know that no $\hat{y} < \epsilon$ can achieve a smaller expected value of term (b) than $\hat{y} = \epsilon$. Combined, we have that
$$
\mathbb{E}_F[\text{TADDA1}^\text{L2}_\epsilon(\epsilon, Y)] \leq \mathbb{E}_F[\text{TADDA1}^\text{L2}_\epsilon(\hat{y}, Y)] \ \text{ for all } \hat{y} < \epsilon.
$$
We can thus restrict our search for the optimal point forecast to $\hat{y} \geq \epsilon$. On this segment of the real line, the expected score is given by 
\begin{align*}
\mathbb{E}_F[\text{TADDA1}^\text{L2}_\epsilon(\hat{y}, Y)] & = \mathbb{E}_F[(\hat{y} - Y)^2] \ \ + \ \ \pi_- \times \mathbb{E}_F[(\hat{y} - \epsilon)^2]\\
& \propto \mathbb{E}_F[(\hat{y} - Z)^2],
\end{align*}
where
$$
Z = \begin{cases}
Y & \text{ with probability } \frac{1}{1 + \pi_-}\\
\epsilon & \text{ with probability } \frac{\pi_-}{1 + \pi_-}.
\end{cases}
$$
The optimal choice for $\hat{y}$ is consequently the mean $\mu_Z$ of $Z$, i.e.,
$$
\hat{y}_\text{OPF} = \mu_Z= \mathbb{E}_F[Z]= \mu \times \frac{1}{1 + \pi_-} + \epsilon \times \frac{\pi_-}{1 + \pi_-}.
$$

\textbf{Case 3 $(\mu < -\epsilon):$} Analogously to Case 2, we have that
$$
\mathbb{E}_F[\text{TADDA1}^\text{L2}_\epsilon(-\epsilon, Y)] \leq \mathbb{E}_F[\text{TADDA1}^\text{L2}_\epsilon(\hat{y}, Y)] \ \text{ for all } \hat{y} > -\epsilon.
$$
Therefore, restricting our search for the optimal point forecast to $\hat{y} \leq - \epsilon$, we have that
\begin{align*}
\mathbb{E}_F[\text{TADDA1}^\text{L2}_\epsilon(\hat{y}, Y)] & = \mathbb{E}_F[(\hat{y} - Y)^2] \ \ + \ \ \pi_+ \times \mathbb{E}_F[(\hat{y} - (-\epsilon))^2]\\
& \propto \mathbb{E}_F[(\hat{y} - Z)^2],
\end{align*}
where
$$
Z = \begin{cases}
Y & \text{ with probability } \frac{1}{1 + \pi_+}\\
-\epsilon & \text{ with probability } \frac{\pi_+}{1 + \pi_+}.
\end{cases}
$$
Again, the optimal choice for $\hat{y}$ is the mean $\mu_Z$ of $Z$, which is given by
$$
\hat{y}_\text{OPF} = \mu_Z=\mathbb{E}_F[Z]= \mu \times \frac{1}{1 + \pi_+} - \epsilon \times \frac{\pi_+}{1 + \pi_+}.
$$

\subsection{Derivation of the optimal point forecast for $\text{TADDA2}^\text{L1}_\epsilon$}

\textit{Result:} The optimal point forecast for 
$$\text{TADDA2}^\text{L1}_\epsilon(\hat{y}, y) = |\hat{y}-y| + a_\epsilon(\hat{y}, y),$$

where $a_\epsilon(\hat{y}, Y)$ is defined as

\begin{equation*}\label{eq:def_tadda2_penalty}
a_\epsilon(\hat{y}, y) = \begin{cases}
    |\hat{y}-\epsilon| & \text{ if } \ \{\hat{y} \le \epsilon \text{ and } y > \epsilon\} \ \text{ or } \ \{\hat{y} > \epsilon  \text{ and } y \in [-\epsilon, \epsilon]\} \\
    |\hat{y}+\epsilon| & \text{ if } \ \{\hat{y} \ge -\epsilon \text{ and } y < -\epsilon\} \ \text{ or } \ \{\hat{y} < -\epsilon \text{ and } y \in [-\epsilon, \epsilon]\}\\
    0 & \text{ otherwise, } 
\end{cases}
\end{equation*}
is given by
$$
\hat{y}_\text{OPF}=
\begin{cases}
F^{-1}\{0.5\times (2 - \pi_-)\} & \text{ if } \pi_- \ge \frac{2}{3}\\
-\epsilon & \text{ if } m<-\epsilon \text{ and } \pi_- < \frac{2}{3} \\
    & \ \ \ \text{ or } -\epsilon \le m \le \epsilon \text{ and } \pi_- \ge \frac{1 + \pi_+ - 2\text{Pr}_F(Y=-\epsilon)}{3} \\
F^{-1}\{0.5\times(1 - \pi_- + \pi_+)\} & \text{ if } -\epsilon \le m \le \epsilon \text{ and } \pi_- < \frac{1 + \pi_+ - 2\text{Pr}_F(Y=-\epsilon)}{3} \ \text{and} \ \pi_+ < \frac{1 + \pi_-}{3} \\
\epsilon & \text{ if } -\epsilon \le m \le \epsilon \text{ and } \pi_+ \ge \frac{1 + \pi_-}{3} \\
    & \ \ \ \text{ or } m>\epsilon \text{ and } \pi_+ \leq \frac{2}{3}\\
F^{-1}(0.5\times \pi_+) & \text{ if } \pi_+ > \frac{2}{3}.
\end{cases}
$$
\\

\textit{Proof:}
The expectation of $\text{TADDA2}^\text{L1}_\epsilon(\hat{y}, Y)$ is given by
$$
\mathbb{E}_F[\text{TADDA2}^\text{L1}_\epsilon(\hat{y}, Y)] = \underbrace{\mathbb{E}_F|\hat{y} - Y|}_{\text{(a)}} \ \ + \ \ \underbrace{\mathbb{E}_F[a_\epsilon(\hat{y}, Y)]}_{\text{(b)}}.
$$
We again proof the result by cases.

\textbf{Case 1} $(-\epsilon \le m \le \epsilon):$
For $\hat{y} = \epsilon$, term (b) is
$$
\mathbb{E}_F[a_\epsilon(\epsilon, Y)] = \pi_- \times 2 \epsilon,
$$
while for $\hat{y} > \epsilon$, we get
$$
\mathbb{E}_F[a_\epsilon(\hat{y}, Y)] = \pi_- \times \underbrace{|\hat{y} + \epsilon|}_{\geq 2\epsilon} + \underbrace{(1 - \pi_- - \pi_+) \times |\hat{y} - \epsilon|}_{\geq 0}.
$$
For $\hat{y} = -\epsilon$, term (b) is
$$
\mathbb{E}_F[a_\epsilon(-\epsilon, Y)] = \pi_+ \times 2 \epsilon,
$$
while for $\hat{y} < -\epsilon$, we have
$$
\mathbb{E}_F[a_\epsilon(\hat{y}, Y)] = \pi_+ \times \underbrace{|\hat{y} - \epsilon|}_{\geq 2\epsilon} + \underbrace{(1 - \pi_- - \pi_+) \times |\hat{y} + \epsilon|}_{\geq 0}.
$$

We can thus conclude that term (b) is smaller for $\hat{y} = \pm\epsilon$ than for any $\hat{y} <- \epsilon$ or $\hat{y} > \epsilon$ and restrict our search for the optimal point forecast to $\hat{y} \in [-\epsilon, \epsilon]$. For this segment of the real line, we obtain
\begin{align*}
\mathbb{E}_F[\text{TADDA2}^\text{L1}_\epsilon(\hat{y}, Y)] &= \mathbb{E}_F|\hat{y} - Y|\ \ + \ \ \pi_- \times \mathbb{E}_F|\hat{y} + \epsilon| \ \ + \ \ \pi_+ \times \mathbb{E}_F|\hat{y} - \epsilon| \\
&\propto \mathbb{E}_F|\hat{y}-Z|,
\end{align*}
where
$$
Z = \begin{cases}
Y & \text{ with probability } \frac{1}{1 + \pi_- + \pi_+}\\
-\epsilon & \text{ with probability } \frac{\pi_-}{1 + \pi_- + \pi_+}\\
\epsilon & \text{ with probability } \frac{\pi_+}{1 + \pi_- + \pi_+},
\end{cases}
$$
The term $\mathbb{E}_F|\hat{y}-Z|$ is minimized by the median $m_Z$ of $Z$. The CDF of $Z$ is given by 
$$
G(z) = \begin{cases}
\frac{1}{1 + \pi_- + \pi_+} \times F(z) & \text{ for } z < -\epsilon\\
\frac{1}{1 + \pi_- + \pi_+} \times F(z) + \frac{\pi_-}{1 + \pi_- + \pi_+} & \text{ for } -\epsilon\le z < \epsilon\\
\frac{1}{1 + \pi_- + \pi_+} \times F(z) + \frac{\pi_- + \pi_+}{1 + \pi_- + \pi_+} & \text{ for } z \ge \epsilon.\\
\end{cases}
$$

Setting $F^{-1}(m_Z) = 0.5$, it can be shown that
$$
\hat{y}_\text{OPF} = m_Z = \begin{cases}
-\epsilon & \text{ if } \pi_- \ge \frac{1 + \pi_+ -2\text{Pr}_F(Y=-\epsilon)}{3} \\
F^{-1}\{0.5\times(1 - \pi_- + \pi_+)\} & \text{ if } \pi_- < \frac{1 + \pi_+ -2\text{Pr}_F(Y=-\epsilon)}{3} \ \text{and} \ \pi_+ < \frac{1 + \pi_-}{3} \\
\epsilon & \text{ if } \pi_+ \ge \frac{1 + \pi_-}{3}.\\
\end{cases}
$$
Note that in many practically relevant cases $-2\text{Pr}_F(Y=-\epsilon) = 0$ holds, so that this term can be omitted in the above equation.

\textbf{Case 2} $(m > \epsilon):$ For $\hat{y} = \epsilon$, term (b) is
$$
\mathbb{E}_F[a_\epsilon(\epsilon, Y)] =  \pi_- \times 2 \epsilon,
$$
while for $-\epsilon \leq \hat{y} < \epsilon$, it is
\begin{align*}
\mathbb{E}_F[a_\epsilon(\hat{y}, Y)] & = \pi_- \times (\hat{y} + \epsilon) + \pi_+ \times (\epsilon - \hat{y})\\
& = (\pi_- + \pi_+) \times \epsilon + (\pi_- - \pi_+) \times \hat{y}\\
& = 2\times \pi_- \times \epsilon + (\pi_+ - \pi_-)\times \epsilon + (\pi_- - \pi_+) \times \hat{y}\\
& = \pi_-\times 2\epsilon + \underbrace{(\pi_+ - \pi_-)}_{< 0}\times \underbrace{(\epsilon - \hat{y})}_{> 0}\\
& > \pi_- \times 2 \epsilon.
\end{align*}
For $\hat{y} < -\epsilon$, we have
\begin{align*}
\mathbb{E}_F[a_\epsilon(\hat{y}, Y)] & = (1 - \pi_- - \pi_+) \times (- \epsilon - \hat{y}) + \pi_+ \times (\epsilon - \hat{y})\\
& = (-1 + \pi_- + 2\pi_+) \times \epsilon + (-1 + \pi_-) \times \hat{y}\\
& = 2\underbrace{\pi_+}_{> \pi_-} \times \epsilon + \underbrace{(-1 + \pi_-)}_{< 0} \times \underbrace{(\hat{y} + \epsilon)}_{< 0}\\
& > \pi_- \times 2 \epsilon.
\end{align*}
We can thus conclude that term (b) is smaller for $\hat{y} = \epsilon$ than for any $\hat{y} < \epsilon$. As we assumed that term (a) is minimized by $m > \epsilon$, we can restrict our search for the optimal point forecast to $\hat{y} \ge \epsilon$ and we have per definition
\begin{align*}
\begin{split}
\mathbb{E}_F[\text{TADDA2}^\text{L1}_\epsilon(\hat{y}, Y)]  = {} & \mathbb{E}_F  |\hat{y} - Y| \ \ + \ \ \pi_- \times \mathbb{E}_F|\hat{y} + \epsilon| \ \ + \ \  (1 - \pi_- - \pi_+) \times \mathbb{E}_F|\hat{y} - \epsilon|
\end{split} \\
 \propto {} & \mathbb{E}_F|\hat{y} - Z|,
\end{align*}
where 
$$
Z = \begin{cases}
Y & \text{ with probability } \frac{1}{2 - \pi_+} \\
- \epsilon & \text{ with probability } \frac{\pi_-}{2 - \pi_+}\\
\epsilon & \text{ with probability } \frac{1 - \pi_- - \pi_+}{2 - \pi_+}.
\end{cases}
$$
$\mathbb{E}_F|\hat{y} - Z|$ is minimized by the median $m_Z$ of $Z$ under $F$. The CDF $G$ of $Z$ can be expressed through $F$ as
\begin{equation}
G(z) = \begin{cases}
\frac{1}{2 - \pi_+} \times F(z) & \text{ if } z < -\epsilon\\
\frac{1}{2 - \pi_+} \times F(z) + \frac{\pi_-}{2 - \pi_+} & \text{ if } -\epsilon \le z < \epsilon\\
\frac{1}{2 - \pi_+} \times F(z) + \frac{1 - \pi_+}{2 - \pi_+} & \text{ if } z \geq \epsilon.
\end{cases}
\label{eq:G_TADDA2_L1_greater}
\end{equation}
We obtain the median $m_Z$ by setting $G(m_Z) = 0.5$. Using
$$
\frac{1}{2 - \pi_+} \times \underbrace{F(\epsilon)}_{= 1 - \pi_+} + \ \  \frac{1 - \pi_+}{2 - \pi_+} \ \geq \  0.5
$$
and some simple algebra, we find that the median is $\epsilon$ whenever $\pi_+ \leq \frac{2}{3}$. This leads to the distinction
$$
\hat{y}_\text{OPF} = m_Z = \begin{cases}
\epsilon & \text{ if } \pi_+ \leq \frac{2}{3}\\
F^{-1}(0.5\times \pi_+) & \text{ otherwise}.
\end{cases}
$$

\textbf{Case 3 $(m < -\epsilon): $}
Following essentially the same arguments as in Case 2, it can be shown that term (b) is smaller for $\hat{y} = -\epsilon$ than for any $\hat{y} > -\epsilon$. As we know that term (a) is minimized by $m <- \epsilon$, we can restrict our search for the optimal point forecast to $\hat{y} \le -\epsilon$. For this part of the real line, we have
\begin{align*}
\begin{split}
\mathbb{E}_F[\text{TADDA2}^\text{L1}_\epsilon(\hat{y}, Y)]  = {} & \mathbb{E}_F |\hat{y} - Y| \ \ + \ \ (1 - \pi_- - \pi_+) \times \mathbb{E}_F|\hat{y} + \epsilon| \ \ + \ \  \pi_+ \times \mathbb{E}_F|\hat{y} - \epsilon|
\end{split} \\
 \propto {} & \mathbb{E}_F|\hat{y} - Z|,
\end{align*}
where 
$$
Z = \begin{cases}
Y & \text{ with probability } \frac{1}{2 - \pi_-} \\
- \epsilon & \text{ with probability } \frac{1 - \pi_- - \pi_+}{2 - \pi_-}\\
\epsilon & \text{ with probability } \frac{\pi_+}{2 - \pi_-}.
\end{cases}
$$
The expectation $\mathbb{E}_F|\hat{y} - Z|$ is minimized by the median $m_Z$ of $Z$. The CDF of $Z$ is given by
\begin{equation*}
G(z) = \begin{cases}
\frac{1}{2 - \pi_-} \times F(z) & \text{ if } z < -\epsilon\\
\frac{1}{2 - \pi_-} \times F(z) + \frac{1 - \pi_- - \pi_+}{2 - \pi_-} & \text{ if } -\epsilon \le z < \epsilon\\
\frac{1}{2 - \pi_-} \times F(z) +\frac{1 - \pi_-}{2 - \pi_-} & \text{ if } z \geq \epsilon.
\end{cases}
\label{eq:G_TADDA2_L1_smaller}
\end{equation*}
From $m < -\epsilon$, we know that $\pi_->0.5$. Using $G(m_Z)=0.5$ and $\hat{y}\le-\epsilon$, we get
$$
\hat{y}_\text{OPF} = m_Z = \begin{cases}
-\epsilon & \text{ if } \pi_- < \frac{2}{3}\\
F^{-1}\{0.5\times (2 - \pi_-)\} & \text{ otherwise}.
\end{cases}
$$
This concludes the proof.

\end{document}